%% file: GAMA_ANN.tex
\documentclass[11pt,a4paper]{article}

\usepackage[T1]{fontenc}
\usepackage[utf8]{inputenc} 
\usepackage{lmodern}

\usepackage{geometry}
\usepackage{graphicx}
\usepackage{amsmath,amssymb}
\usepackage{bm}
\usepackage{physics}

\usepackage{natbib}
\usepackage{hyperref}
\usepackage{xcolor}

\hypersetup{
  colorlinks=true,
  linkcolor=blue,
  citecolor=blue,
  urlcolor=blue
}

\def\shark{\textsc{Shark}}

\title{A Demonstration of a Neural Network as a Bridge Between Galaxy Simulations and Surveys}

\author{
E. Elson$^{1}$}

\date{
$^{1}$Department of Physics \& Astronomy, University of the Western Cape, \\Robert Sobukwe Road, Bellville 7535, South Africa\\
\today
}

\input{journals}

\begin{document}
\maketitle

\begin{abstract}
This paper demonstrates that the stellar masses of galaxies in the Galaxy and Mass Assembly (GAMA) survey, originally derived via stellar population synthesis modelling, can be accurately predicted using only their absolute magnitudes and colour indices. A central contribution of this work is the demonstration that this long-standing inference problem can be solved using an exceptionally simple machine-learning model: a fully connected, feed-forward artificial neural network with a single hidden layer.  The network is trained exclusively on synthetic galaxies generated by the \textsc{Shark} semi-analytic model and is shown to transfer effectively to real observations. Across nearly 3.5~dex in stellar mass, the predicted values closely track the GAMA SED-derived masses, with a typical scatter of $\sim0.131$~dex. These results demonstrate that complex deep-learning architectures are not a prerequisite for robust stellar mass estimation, and that simulation-trained, lightweight machine-learning models can capture the dominant physical information encoded in broad-band photometry. The method is further applied to 17\,006 GAMA galaxies lacking SED-derived masses, with photometric uncertainties propagated through the network to provide corresponding error estimates on the inferred stellar masses. Overall, this work establishes a computationally efficient and conceptually transparent pathway for simulation-to-observation transfer learning in galaxy evolution studies.
\end{abstract}

\input{intro}
\input{data_ANN}
\input{results}
\input{no_mstar}

\input{summary}

\bibliographystyle{plainnat}

\end{document}

%% file: journals.tex
\newcommand{\apj}{ApJ}
\newcommand{\apjl}{ApJL}

\newcommand{\mnras}{MNRAS}
\newcommand{\pasp}{PASP}
\newcommand{\araa}{ARA\&A}

\newcommand{\apss}{Ap\&SS}

%% file: intro.tex
\section{Introduction}

Estimating the stellar masses of galaxies remains a central challenge in extragalactic astronomy. Stellar mass is not a direct observable; instead, it is typically inferred through spectral energy distribution (SED) fitting, which requires assumptions about star formation histories, metallicities, dust attenuation, and the stellar initial mass function, among other factors \citep{pforr_2012, conroy_2013}. These modelling choices can vary widely across studies, introducing systematic uncertainties and rendering stellar mass estimates inherently model-dependent \citep{courteau_2014}.

In contrast, cosmological galaxy formation simulations predict stellar mass as a fundamental outcome, with models calibrated to reproduce observed stellar mass functions (e.g., \citealt{schaye_2015, pillepich_2018, dave_2019}). Such simulations provide physically motivated predictions for galaxy growth, colours, and spectral properties, naturally encoding relationships between photometry and stellar mass. An emerging question in modern galaxy evolution work is therefore whether theoretical information from simulations can be transferred efficiently into observational domains to aid or accelerate stellar mass estimation. The difficulty lies in bridging the gap between simulated and real galaxies, which need not populate the same regions of observable parameter space.

Machine learning offers one route across this divide. In particular, feed-forward artificial neural networks (ANNs) are capable of learning complex, non-linear mappings between photometric inputs and physical galaxy properties. In \citet{elson_2025}, it was shown that even an extremely simple network, consisting of a fully connected feed-forward ANN with a single hidden layer, trained exclusively on synthetic galaxies generated by the \textsc{Shark} semi-analytic model \citep{lagos_2018}, was able to recover simulated stellar masses with high fidelity using only broadband photometry and colours. That work demonstrated not only that simulated galaxies contain enough observable information to predict stellar mass, but also that sophisticated architectures (e.g. deep CNNs or encoder–decoder frameworks) are not required for this task. Crucially, it was suggested that such a model might be transferable to real data.

The present study therefore focuses not on developing a new model, but on testing whether the previously trained ANN from \citet{elson_2025} can be applied directly to real observational data. Aside from minor adjustments to input features where observational coverage differs, the architecture and training set (of simulated galaxy properties) are unchanged. The central aim is to evaluate whether a simulation-trained network can recover stellar masses for real galaxies in the Galaxy And Mass Assembly (GAMA) survey \citep{GAMA}, and to assess its performance by comparison with SED-derived masses and through an application to galaxies lacking SED-based estimates. In this way, the study provides a direct test of simulation-to-observation transfer learning for stellar mass inference.

%% file: data_ANN.tex
\section{GAMA data and the Artificial Neural Network}

Stellar masses for GAMA galaxies were taken from the \texttt{StellarMassesLambdarv24} table, available via the GAMA DR4 Schema Browser\footnote{\url{https://www.gama-survey.org/dr4/schema/table.php?id=691}}. These masses were derived from broad-band SED fitting \citep{taylor_2011}, using optical SDSS (DR7) and near-infrared UKIDSS LAS (DR4) photometry. The observed SEDs were compared to the stellar population models of \citet{bruzual_2003}, spanning a range of star formation histories, metallicities, and dust attenuations, and assuming a \citet{chabrier_2003} initial mass function. Bayesian fitting of the stellar mass-to-light ratios produced stellar masses with typical uncertainties of $\lesssim 0.2$~dex.

In \citet{elson_2025}, it was demonstrated that the absolute magnitudes of galaxies across the far-ultraviolet to far-infrared range, as generated by the \shark\ semi-analytic model, could be used as input features to a simple feed-forward neural network to predict stellar masses.  The present study applies the same trained model to observational data. Specifically, the measured absolute magnitudes and colour indices of GAMA galaxies are provided as inputs to the network, with the aim of recovering their stellar masses as derived from SED fitting.

Ideally, the same feature set used in \citet{elson_2025} would be available for GAMA galaxies so that the network could be applied directly. In practice, two limitations arise. First, the GAMA \texttt{gkvScienceCatv02} table does not provide Spitzer $W3$ and $W4$ magnitudes, which also removes $W2-W3$ as a viable colour index. Second, for the network to operate reliably, the distribution of each GAMA feature must lie within the range spanned by the corresponding \shark\ training data. All available GAMA features satisfy this condition except for $FUV-NUV$ and $W1-W2$, which lie far outside the training ranges and are therefore excluded. For a small number of other features, only a minor fraction of GAMA galaxies extended slightly beyond the \shark\ limits. In these cases, affected galaxies were removed so that the input set remained fully representative of the simulation feature space. 

\begin{figure*}
    \centering
    \includegraphics[width=\linewidth]{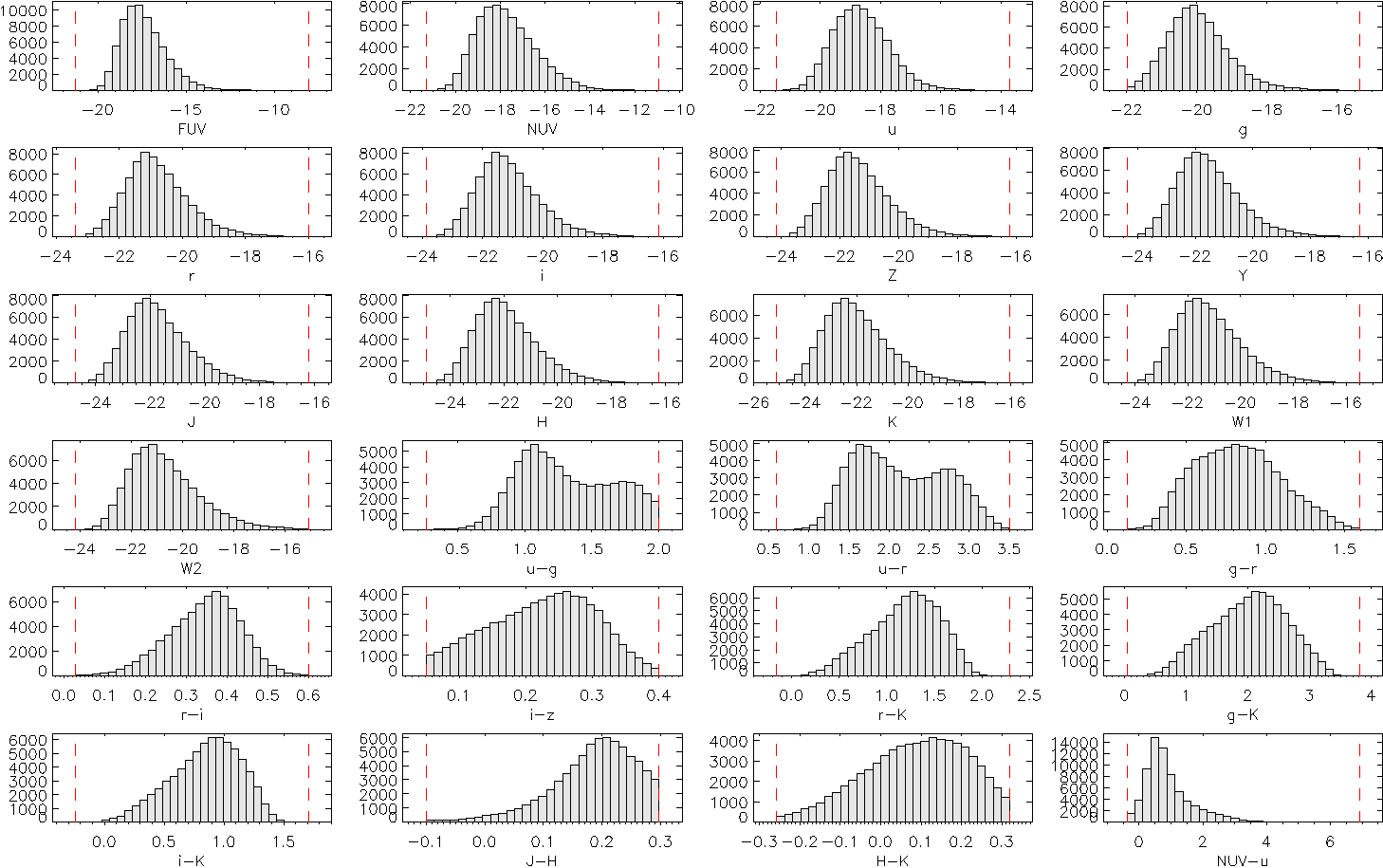}
\caption{
Distributions of the 24 broadband absolute magnitudes and colour indices (all in the AB system) used as input features for the GAMA galaxies in this study. In each panel, the red dashed vertical lines indicate the minimum and maximum values spanned by the corresponding features in the \shark\ training sample. GAMA galaxies lying outside these ranges were excluded to ensure that all inputs fall within the domain on which the ANN was trained.
}
    \label{GAMA_props}
\end{figure*}

After cross-matching GAMA galaxies with photometry in \texttt{gkvScienceCatv02} and stellar masses in \texttt{StellarMassesLambdarv24}, a final sample of 71\,171 galaxies was obtained. Only galaxies with $NQ>2$ (normalised redshift quality, ensuring robust spectroscopic redshifts) and $SC>3$ (science sample class, restricting to unique galaxies within the main $r<20.5$ survey selection in reliable regions of the sky) were retained. The distributions of magnitudes and colours are shown in Fig.~\ref{GAMA_props}. Apparent magnitudes were converted to absolute magnitudes using the luminosity distance, $D_\mathrm{L}=(1+z)D_\mathrm{comov}$, where $z$ is the spectroscopic redshift and $D_\mathrm{comov}$ is the comoving distance.

Because the $W3$, $W4$, $W2-W3$, $W1-W2$ and $FUV-NUV$ features are unavailable or incompatible, the \citet{elson_2025} network was retrained using the available 24 magnitudes\footnote{GALEX $FUV$ and $NUV$; SDSS $u$, $g$, $r$, $i$, $z$; VISTA $Y$, $J$, $H$, $Ks$; and Spitzer $W1$, $W2$.} and colours\footnote{$u-g$, $u-r$, $g-r$, $r-i$, $i-z$, $r-Ks$, $g-Ks$, $i-Ks$, $J-H$, $H-Ks$, $NUV-u$.}. In addition, the bulge-to-total mass ratio cut ($B/T<0.65$) employed in that study to remove massive, AGN-dominated systems was removed here to ensure coverage up to $\sim 10^{11.5}~M_\odot$. With these adjustments, the retrained ANN achieves an error of $\sim 0.117$~dex on \shark\ galaxies, compared to $0.087$~dex reported by \citet{elson_2025}. The reduced performance is expected, particularly since $FUV-NUV$ was shown to be a powerful diagnostic colour through which the original ANN implicitly learned ultraviolet attenuation behaviour.

%% file: results.tex
\section{Results}
\begin{figure}
    \centering
    \includegraphics[width=0.7\linewidth]{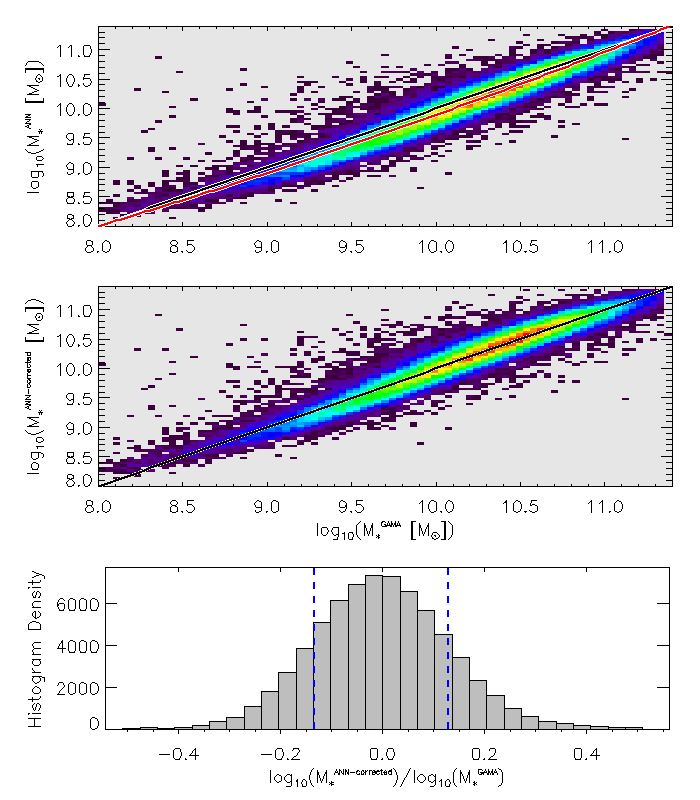}
    \caption{
\textbf{Top:} Raw ANN-predicted stellar masses compared with SED-derived masses for 71\,171 GAMA galaxies. The solid line denotes the one-to-one relation, while the red curve shows a polynomial fit to the data. 
\textbf{Middle:} Corrected ANN predictions after subtracting the fitted residual trend, removing the median offset and bringing the relation into alignment with the one-to-one line. 
\textbf{Bottom:} Distribution of corrected residuals, with half of the 16--84 percentile range (indicated by the blue-dashed lines) being 0.135~dex. Together, the panels demonstrate that the ANN closely reproduces SED-derived stellar masses with a small correctable bias and low overall scatter.
}
    \label{results}
\end{figure}

The top panel of Fig.~\ref{results} shows the stellar masses predicted by the ANN as a function of those derived via broad-band SED fitting. The solid black line indicates the one-to-one relation, while the red curve is a second-order polynomial fit to all 71\,171 individual galaxy measurements (not to the binned 2D distribution). It is immediately apparent that the direct application of an ANN trained solely on synthetic \shark\ galaxies performs remarkably well on real GAMA data. Over the full mass range probed, the predicted stellar masses follow the SED-derived values closely.  The small, almost constant offset of up to $\sim 0.1$~dex likely arises because the ANN learns the mass-photometry relationship encoded in \shark, whereas the GAMA stellar masses reflect assumptions embedded in SPS modelling.  The two frameworks need not map colours to stellar mass in exactly the same way.  The predictive scatter is quantified using half of the 16--84 percentile range of the residual distribution (a robust estimator of the 1$\sigma$ scatter for near-Gaussian data), yielding a value of 0.135~dex.

The median discrepancy can be quantified by taking the difference between the black one-to-one line and the red polynomial fit in the top panel. This residual reaches a maximum of $\sim -0.11$~dex at $M_*\approx 10^{9.7}~M_{\odot}$, slightly below the median GAMA stellar mass of $\approx 10^{10.2}~M_{\odot}$. This smooth residual function can be used to correct any ANN-predicted  stellar mass. Applying this correction to all predictions removes the systematic offset.. The second panel of Fig.~\ref{results} shows the corrected ANN masses plotted against the SED-derived masses, where the median alignment with the one-to-one relation is restored. The final panel shows the distribution of residuals between the corrected ANN predictions and the SED-derived stellar masses.  Note that  the scatter from the original comparison is retained as expected for a simple regression bias correction.

These results show that an artificial neural network trained solely on simulated galaxy photometry can, with minimal adjustments, reproduce the stellar masses of real galaxies with high fidelity. Although the model was never exposed to observational data during training, it generalises effectively, accurately matching SED-derived stellar masses. This provides a promising proof of concept for future simulation-to-observation transfer learning in galaxy evolution studies.

%% file: no_mstar.tex
\section{Application to GAMA Galaxies Without SED-derived Masses}

As an additional demonstration, the trained ANN was applied to GAMA galaxies which lack SED-derived stellar masses. These 17\,006 objects were identified by selecting sources with $SC>4$ in the \texttt{gkvScienceCatv02} catalogue that do not appear in \texttt{StellarMassesLambdarv24}. For each galaxy, the input fluxes required by the ANN were extracted along with their quoted photometric uncertainties. To propagate these errors into the predicted stellar masses, each flux $f$ was perturbed by half of its associated uncertainty, generating two realisations $f+\delta f/2$ and $f-\delta f/2$, where $\delta f$ is the reported flux error. The ANN was evaluated three times for each galaxy using $\{f-\delta f/2,\,f,\,f+\delta f/2\}$, providing both a central mass estimate and corresponding upper and lower bounds.

\begin{figure}
    \centering
    \includegraphics[width=0.7\linewidth]{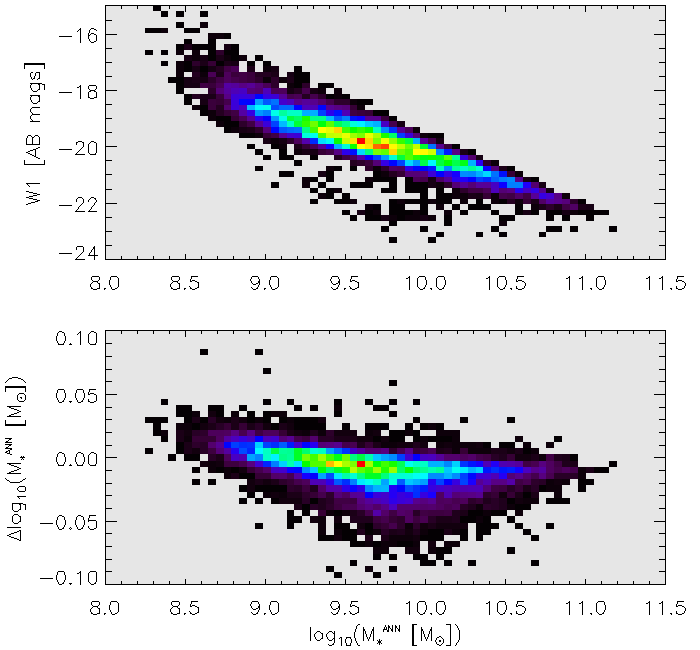}

\caption{
\textbf{Top:} WISE $W1$ magnitude vs.\ ANN-predicted stellar mass for GAMA galaxies without SED estimates, showing a clear linear relation consistent with $W1$ tracing stellar mass. 
\textbf{Bottom:} Propagated uncertainties on $\log_{10}(M_\star)$ as a function of predicted mass, typically $\pm0.05$~dex and largest at intermediate masses, where galaxy populations are most diverse in dust content and star-formation history.
}
    \label{fig:predicted_noSED}
\end{figure}

The predicted stellar masses for this sample are shown in the top panel of Fig.~\ref{fig:predicted_noSED}. When plotted against the WISE $W1$ magnitude, a clear and strongly linear relationship emerges in log--log space, as expected given that $W1$ flux is known to trace stellar mass \citep{jarrett_2023}. This behaviour provides confidence that the ANN is producing physically reasonable stellar mass estimates even for galaxies without SED-derived values.

The lower panel of Fig.~\ref{fig:predicted_noSED} displays the propagated uncertainty on $\log_{10}(M_*)$ as a function of the predicted mass. The uncertainties lie in the range $\sim{-0.05}$ to $+0.05$~dex and exhibit a characteristic peak near the middle of the mass range. This behaviour is physically expected: intermediate-mass galaxies span a wide variety of star-formation histories, dust contents, and quenching states, making their photometric mass-to-light ratios more degenerate than those of low-mass star-forming systems or high-mass quiescent systems.

The flux-propagated uncertainties are significantly smaller than the intrinsic ANN scatter of $\sim0.131$~dex measured earlier for galaxies with known SED masses. Since the intrinsic scatter already incorporates observational noise, modelling uncertainties, and simulation-to-observation transfer limitations, the flux-driven uncertainties should not be added in quadrature. Instead, the $\sim0.05$~dex perturbation range reflects only the sensitivity of predictions to flux measurement error, while the $\sim0.131$~dex scatter remains the dominant uncertainty component. A very conservative uncertainty on ANN-predicted stellar masses is therefore $0.131+0.05\sim0.18$~dex.

For context, it is useful to compare the ANN-based stellar mass uncertainties with those obtained using more traditional methods. Stellar masses derived from SED modelling typically carry uncertainties of $\sim0.2$--$0.3$~dex due to degeneracies among stellar age, metallicity, dust attenuation, and star-formation history \citep[e.g.,][]{walcher_2011, conroy_2013}. Moreover, comparisons between different SED-based mass estimates often reveal non-negligible systematic offsets \citep[e.g.,][]{elson_2024}, emphasising that such masses are themselves model-dependent.

Empirical scaling relations provide alternative estimates but are likewise limited by intrinsic scatter.  The baryonic Tully--Fisher relation, for example, yields uncertainties of $\sim0.1$--$0.4$~dex \citep{bradford_2016, lelli_2016}. In this context, the ANN-based approach presented here achieves a level of performance comparable to commonly used methods while requiring only broadband photometry. Even adopting a conservative uncertainty estimate of $\sim0.18$~dex, the results indicate that simulation-trained neural networks provide a practical and complementary tool for stellar mass estimation in large galaxy surveys.

%% file: summary.tex
\section{Summary}


This study demonstrates that galaxy stellar masses can be accurately inferred using an exceptionally simple machine-learning model trained solely on simulations. In particular, a fully connected, feed-forward artificial neural network with only a single hidden layer-trained exclusively on synthetic galaxies from the \textsc{Shark} semi-analytic model-is shown to perform robust stellar mass estimation for real galaxies in the GAMA survey using only broadband magnitudes and colour indices as input features. To the author's knowledge, no existing work has demonstrated comparable performance over such a wide dynamic range using a neural architecture of this minimal complexity.

After minor adjustments to ensure compatibility between simulated and observed feature ranges, the network successfully reproduces SED-derived stellar masses across approximately 3.5 dex, with a typical scatter of $\lesssim 0.131$ dex. This result directly demonstrates that complex neural architectures are not a prerequisite for robust stellar mass inference, and that the dominant physical information required for this task is already encoded in broad-band photometry. The model is further applied to 17\,006 GAMA galaxies lacking SED-based mass estimates, with photometric uncertainties propagated through the network to yield typical stellar mass uncertainties of only $\sim 0.05$ dex. Overall, these results establish a computationally efficient and conceptually transparent simulation-to-observation transfer learning framework, enabling rapid stellar mass estimation for large galaxy samples and substantially lowering the barrier to applying machine-learning techniques in galaxy evolution studies.